\documentclass[12pt,a4paper]{article} 

\usepackage{feynmf,amsmath,amssymb,latexsym,epsfig,floatflt,subfigure}
\def\sst{\scriptscriptstyle}

\newcommand{\be}{\begin{equation}}
\newcommand{\ee}{\end{equation}}
\def\bea{\begin{eqnarray}}
\def\eea{\end{eqnarray}}

\def\NPB#1#2#3{Nucl. Phys. {\bf B} {\bf#1} (#2) #3}
\def\PLB#1#2#3{Phys. Lett. {\bf B} {\bf#1} (#2) #3}
\def\PRD#1#2#3{Phys. Rev. {\bf D} {\bf#1} (#2) #3}
\def\PRL#1#2#3{Phys. Rev. Lett. {\bf#1} (#2) #3}

\def\NCA#1#2#3{Nuovo Cim. A {\bf#1} (#2) #3}

\def\HEP#1{arXiv:hep-ph/#1}
\def\HEPEX#1{arXiv:hep-exp/#1}
\def\JMP#1#2#3{J. Math. Phys {\bf#1}(#2)#3}
\def\JHEP#1#2#3{JHEP {\bf#1} (#2) #3}
\begin{document} 
\thispagestyle{empty}
\begin{titlepage}

\title{\bf Spontaneous CP Violation in SUSY SO(10)}

\author{Yoav Achiman\footnote{e-mail:achiman@post.tau.ac.il}\\[0.5cm]                   
  School of Physics and Astronomy\\
  Tel Aviv University\\
  69978 Tel Aviv, Israel\\[1.5cm]}

\date{ March  2007}
 
\maketitle
\setlength{\unitlength}{1cm}
\begin{picture}(5,1)(-12.5,-12)
\put(0,0){TAUP 2852/07}
\end{picture}
\parindent 0cm
\begin{abstract}
\noindent
A scenario is suggested for spontaneous $CP$ violation in non-$SUSY$ and
$SUSY\ SO(10)$. The idea is to have a scalar potential which generates 
spontaneously a phase, at the high scale, in the VEV that gives a mass
to the $RH$ neutrinos. As a possible realization 
the case of the minimal renormalizable $SUSY\ SO(10)$ is discussed in detail 
% It is demonstrated that this induces also a phase in the $CKM$ matrix.
and one finds that a phase is induced in the $CKM$ matrix.
It is also pointed out that, in these models, the scales of Baryogenesis, 
Seesaw, Spontaneous $CP$ violation and Spontaneous $U(1)_{PQ}$ breaking are 
all of the same order of magnitude. 
\end{abstract}
\thispagestyle{empty} 
\end{titlepage}
 
%\setlength{\parindent}{0pt} 
%\vskip 1cm
\parindent 0pt 

%------------------------------  1  ------------------------------------------

\parindent=0pt
%\begin{document}

There are three manifestations of CP violation in Nature:\\

1) {\em Fermi scale CP violation} as is observed in the $K$ and $B$ decays 
\cite{KB}. This violation is induced predominantly by a complex mixing matrix
 of the quarks ($CKM$).\\

2) {\em The cosmological matter antimatter asymmetry ($BAU$)}
is an indication for high scale $CP$ violation \cite{Sakharov}.
In particular, it's most popular explanation via leptogenesis \cite{Yanagida} 
requires $CP$ breaking decays of the heavy right-handed ($RH$) neutrinos.\\

3) {\em The strong $CP$ problem } called also the $QCD$ $\Theta$ problem
\cite{window} lies in the non-observation of $CP$ breaking in the strong
interactions while there is an observed $CP$ violation in the interaction of
quarks.\\

It is still not clear if there is one origin to those $CP$ breaking 
manifestations.
What is the nature of the violation of $CP$ ? Is it intrinsic in terms of 
complex 
Yukawa couplings or due to spontaneous generation of phases in the Higgs 
VEVs ?\\

{\em Spontaneous violation of $CP$} \cite{T.D.Lee} is more difficult to realize
but has advantages with respect to the intrinsic ones:\\

1) It is more elegant and involves less parameters. The intrinsic breaking 
becomes quite arbitrary in the framework of $SUSY$ and $GUT$ theories.\\

2) It solves the $SUSY$ $CP$ violation problem (too many potentially complex 
parameters) as all parameters are real.\\

3) It leads to the vanishing of $\Theta_{QCD}$ (but not ArgDetM) at the tree 
level. This can be used as a first step towards solving the $CP$ problem 
by adding extra symmetries and exotic quarks~\cite{Nelson}\cite{Barr}
\cite{branco}.\\

For good recent discussion of spontaneous $CP$ violation ($SCPV$), with many 
references, see Branco and Mohapatra~\cite{BM}.\\

It is preferable to break $CP$ at a high scale. This is what we need for
the $BAU$. Especially if this is due to leptogenesis  i.e.
$CP$ violating decays of heavy neutrinos, it is mandatory. This is also needed
to cure the domain wall problem~\cite{Zeldovich}.\\ 
Also, $SCPV$ cannot take place in the standard model ($SM$) because of gauge 
invariance. Additional Higgs bosons must be considered and those lead 
generally to flavor changing neutral currents. The best way to avoid these is 
to make the additional scalars heavy~\cite{BM}.\\
In this case, the scale of $CP$ violation can be related to the seesaw 
\cite{seesaw} scale as well
as to the $U(1)_{PQ}$ \cite{PQ} breaking scale, i.e. the ``axion window''
\cite{window}.\\

%-------------------------- 2  ------------------------------------------
\pagebreak

Branco, Parada and Rebelo discussed in their paper~\cite{branco} also the
possibility of a common origin to all $CP$ violations. Their model however is
in the framework of a non-SUSY Standard Model (SM), extended by a heavy complex
singlet Higgs and an exotic vectorlike quark.\\

I would like to suggest in this letter a scenario, along this line~\cite{YA}
\cite{pap}, for $SVCP$ in SUSY GUTs by giving an explicit realization in the 
framework of the {\em minimal renormalizable SUSY SO(10)} \cite{minimal} 
(without the need for exotic fermions).\\

As an introduction 
let me start by revising the {\em renormalizable non-$SUSY$ $SO(10)$} 
and a possible $SCPV$~\cite{YA}.\\ 

non-SUSY GUTs require  intermediate  gauge symmetry 
breaking ($I_i$) \cite{inter} to have gauge coupling unification.
\be
GUT \longrightarrow I_i \longrightarrow SM = SU_C(3)\times
SU_L(2)\times U_Y(1)\ .
\ee
Most  models involve an intermediate scale at $\approx 
10^{12-13} GeV$ which is also that of  breaking of $B-L$, the masses of $RH$ neutrinos
and the $CP$ violation~responsible~for~leptogenesis ($BAU$).\\

$SO(10)$ fermions are in three ${\bf 16}$ representations: $\Psi_i(\bf 16)$.
\be
{\bf 16}\times{\bf 16} = ({\bf 10} + {\bf 126})_S
+ {\bf 120}_{AS}\ .
\ee

Hence, only $H({\bf 10)},\ {\overline\Sigma}(\overline{\bf 126})$\ and\
$D({\bf 120})$ can contribute directly to Yukawa couplings and 
fermion masses. Additional Higgs representations are needed for the 
gauge symmetry breaking. \\
One and only one $VEV$ \quad
${\overline\Delta} = <{\overline\Sigma} (1,1,0) >$\quad
can give a (large) mass to the $RH$ neutrinos via
\be
Y_\ell ^{ij} \nu_{\scriptscriptstyle{R}}^i {\overline\Delta}
 \nu_{\scriptscriptstyle{R}}^j 
\ee
and so induces the seesaw mechanism.
It breaks also \quad {$B-L$}\quad and $SO(10)\rightarrow 
SU(5)$.\\

To generate $SCPV$ in conventional SO(10) one can use the fact that
${\overline\Sigma}({\overline {126}})$ is the only relevant complex Higgs 
representation. Its other special property is that $({\overline\Sigma})_
{\sst{S}}^{\sst{4}}$
is invariant in $SO(10)$~\cite{HHR}. This allows for a $SCPV$ at the high
scale, using the scalar potential~\cite{YA}:
\be
V = V_0 + \lambda_1(H)_{\sst{S}}^{\sst{2}} [({\overline\Sigma})_
{\sst{S}}^{\sst{2} }
+ ({\overline\Sigma}^*)_{\sst{S}}^{\sst{2}}] +
\lambda_2[({\overline\Sigma})_{\sst{S}}^{\sst{4}} + 
({\overline\Sigma}^*)_{\sst{S}}^{\sst{4}}]\ .
\ee
Inserting the $VEV$s

\be
<H(1,2,-1/2)> = \frac{v}{\sqrt{\sst{2}}}
\ \ \ \ \  \overline\Delta = \frac{\sigma}{\sqrt{\sst{2}}}
{e^{i\alpha}}
\ee

in the neutral components, the scalar potential reads
\be
V(v,\sigma,\alpha) = A\cos (2\alpha) + B\cos (4\alpha)\ .
\ee

%------------------------------ 3  --------------------------------
\pagebreak

For $B$ positive and $|A| > 4B$ the absolute minimum of the 
potential requires
\be
\alpha = \frac{1}{2}\arccos\left( \frac{A}{4B}\right ). 
\ee
This ensures  the spontaneous breaking of $CP$ \cite{branco}.\\

It is not possible to realize the above scenario in renormalizable SUSY 
theories, as $\Phi^4$ cannot be generated from the superpotential in this 
case. A different approach is needed and this is the aim of this paper.\\

I will present in the following a possible scenario for $SCPV$ in 
renormalizable SUSY SO(10) models \cite{maryland} \cite{trieste} 
\cite{japan} \cite{wuppertal}. 
This will be done by giving an explicit realization in terms of the so called
{\em the minimal renormalizable SUSY SO(10) model} \cite{minimal}. The model
became very popular recently
%~\cite{maryland}~\cite{trieste}~\cite{japan}~\cite{wuppertal}
due to its simplicity, predictability and automatic $R$-parity invariance
(i.e. a dark matter candidate).\\

It includes the following Higgs representations
\be
H({\bf 10}),\quad \Phi ({\bf 210}), \quad\Sigma ({\bf 126}) \oplus {\overline
\Sigma} ({\overline{\bf 126}})\ .
\ee
Both $\Sigma$ and $\overline{\Sigma}$ are required to avoid high scale $SUSY$ 
breaking ($D$-flatness) and $\Phi ({\bf 210})$ is  needed for the gauge breaking.
 \\

The properties of the model are dictated by the superpotential. This involves
all possible renormalizable products of the superfields

\be
W=M_\Phi \Phi^2 + \lambda_\Phi \Phi^3 + M_\Sigma \Sigma\overline\Sigma
+ \lambda_\Sigma \Phi\Sigma\overline\Sigma
+ M_{\scriptscriptstyle H} H^2 + \Phi H(\kappa\Sigma + \bar\kappa
\overline\Sigma) + 
\Psi_i(Y_{\scriptscriptstyle\bf 10}^{ij}H 
+ Y_{\overline{\scriptscriptstyle\bf 126}}^{ij} \overline\Sigma)\Psi_j
\ee
(One can, however, add discrete symmetries or $U(1)_{PQ}$ invariance etc. 
on top of $SO(10)$).\\

We take all coupling constants real and positive, also in the soft $SUSY$
breaking terms.\\

The symmetry breaking goes in two steps

\be
SUSY SO(10) \stackrel{strong\ gauge\ breaking}
{\longrightarrow} MSSM 
\stackrel{{\scriptscriptstyle SUSY}\ breaking}{\longrightarrow} SM
\ee

The $F$ and $D$-terms must vanish during the strong gauge 
breaking to avoid high scale $SUSY$ breakdown ("$F$,$D$ 
flatness").\\

{\em $D$-flatness}:\quad
 only $\Sigma$, $\overline\Sigma$ are relevant therefore
\be
|\Delta| = |\bar\Delta| \ \hbox{ i.e. }\ \sigma  = \bar\sigma\ .
\ee
%----------------------------- 4 ---------------------------------
The situation with\quad {\em $F$-flatness}\quad is more complicated.\\ 
The strong breaking
 is dictated by the $VEV$s that are $SM$ singlets.\\ 
Those are in the ${\scriptstyle SU_C(4)\times SU_L(2) \times SU_R(2})$  
notation :

$$
%\sst
\phi_1 = <\Phi(1,1,1)> \ \  \phi_2 = <\Phi(15,1,1)> \ \  
\phi_3 = <\Phi(15,1,3)>
$$
$$
%\sst
\Delta = <\Sigma(\overline{10},1,3)> \ \ \ \  \bar\Delta = 
<\bar\Sigma(10,1,3)> .
$$

The strong breaking superpotential in terms of those $VEV$'s is then
\be
\begin{array}{ccl}
%\sst
W_H & = & M_\phi(\phi_1^2 + 3\phi_2^2 + 6\phi_3^2)
   + 2\lambda_\phi(\phi_1^3 + 3\phi_1\phi_2^2 + 6\phi_2\phi_3^2)\\
[10pt]
&+& M_W\Delta\bar\Delta + \lambda_\Sigma\Delta\bar\Delta
(\phi_1 +3\phi_2 + 6\phi_3) .
\end{array}
\ee
$\sst \left| \frac{\partial W_H}{\partial 
v_i}\right|^2 = 0$ \qquad
gives a set of equations.
Their solutions dictate the details of the strong symmetry breaking.\\

One chooses the parameters such that the breaking 
$$
SUSY SO(10) \longrightarrow MSSM
$$
will be achieved~\cite{Senjan}~\cite{Fukuyama}.
$SUSY$ is broken by the soft $SUSY$ breaking terms.
The gauge $MSSM$ breaking is induced by the $VEV$'s
of the $SM$ doublet $\phi^{u,d}(1,2,\pm 1/2)$ components of the Higgs 
representations.\\

The mass matrices of the Higgs are then as follows
\be
M_{ij}^u = \left[\frac{\partial^2W}{\partial\phi_i^u
\partial\phi_j^u}\right]_{\phi_i=<\phi_i>}\ \ \ 
M_{ij}^d = \left[ \frac{\partial^2W}{\partial\phi_i^d
\partial\phi_j^d}\right]_{\phi_i = <\phi_i>} .
\ee
The requirement 
\be
\det(M_{ij}^u) \approx 0 \ \ \ \det(M_{ij}^d)\approx 0
\ee
leaves only two light combinations of doublet components and those play the 
role of the bidoublets \quad $h_u , h_d$ \quad of the $MSSM$. (This also is 
discussed in detail in the papers of~\cite{Senjan}~\cite{Fukuyama}.) \\

We will come back to $h_u , h_d$ later but let me discuss the $SCPV$ 
first.\\

As in the non-SUSY case, we conjecture that $\Delta$ and $\bar\Delta$, 
and only those, acquire a phase at the tree level
\be
<\Sigma(1,1,0)> \equiv \Delta = \sigma e^{i\alpha}\ \ \ \ 
<\bar\Sigma(1,1,0)> \equiv \bar\Delta = \sigma e^{i\bar\alpha} .
\ee

Let me show that this is a minimum of the scalar potential in a
certain region of the parameter space.\\

%--------------------------- 5  --------------------------------------
\pagebreak

To do this we collect all terms with 
$\Delta, \bar\Delta$ in the 
superpotential. Those involve the $VEV$'s that are non-singlets 
under the $SM$. I.e. the $SM$ doublet components of the Higgs 
representations.
\be
\begin{array}{ccccccc}
\phi^u&=&<\Phi(1,2,1/2)>&\ \ \ &\phi^d&=&<\Phi(1,2,-1/2)>\\
H^u &=& <H(1,2,1/2)>&\ \ \ &H^d &=& <H(1,2,-1/2)>\\
\Delta^u&=&<\Sigma(1,2,1/2)>& & \Delta^d&=&<\Sigma(1,2,-1/2)>\\
\bar\Delta^u&=&<\bar\Sigma(1,2,1/2)>& &\bar\Delta^d & = &
<\bar\Sigma(1,2,-1/2)>\\
\end{array}
\ee

The relevant terms are:
\begin{eqnarray}
W_\Delta &=& M_\Sigma \Delta\bar\Delta + 
\frac{\lambda_\Sigma}{10} (\phi^u\Delta^d\bar\Delta +
\phi^d\bar\Delta^u\Delta) \nonumber \\ 
&+& (\frac{\lambda_\Sigma}{10}(\frac{1}{\sqrt{6}}\phi_1\Delta
\bar\Delta + \frac{1}{\sqrt{2}}\phi_2\Delta\bar\Delta
+\phi_3\Delta\bar\Delta )\\
&+& \frac{\lambda_\Sigma\sqrt{2}}{15} \phi_2\bar\Delta^u
\Delta^d - \frac{\kappa}{\sqrt{5}} \phi^d H^u\Delta -
\frac{\bar\kappa}{\sqrt{5}} \phi^u H^d \bar\Delta \nonumber
\end{eqnarray}
using ~\cite{Senjan}~\cite{Fukuyama}.\\

One can then calculate the corresponding scalar potential
\be
V(\alpha,\bar\alpha,M_\Sigma,\lambda_\Sigma,\kappa,\bar\kappa,v_i) =
\sum_i \left| \frac{\partial W_\Sigma}{\partial v_i}
\right|^2\ .
\ee
Noting that  \qquad$|A+Be^{i\alpha} |^2 = A^2 + B^2 +2AB \cos\alpha$\\

and \qquad $|K+P\Delta\bar\Delta|^2 = K^2 + P^2\sigma\bar\sigma +
2KP\sigma\bar\sigma\cos(\alpha+\bar\alpha)$,\\

one finds that\\

$V=A(M_\Sigma, \lambda_\Sigma, \kappa, \bar\kappa, v_i) + 
B(M_\Sigma, \lambda_\Sigma, \kappa, \bar\kappa, v_i)\cos\alpha+$\\

$ D(M_\Sigma, \lambda_\Sigma, \kappa, \bar\kappa ,v_i)
\cos\bar\alpha + E(M_\Sigma, \lambda_\Sigma, \kappa, \bar\kappa,
v_i)\cos(\alpha+\nobreak\bar\alpha)$ .\\

For explicit expressions of the coefficients see the Appendix.\\

The minimalization under $\alpha, \bar\alpha$ requires

\be
\begin{array}{ccc}  
\frac{\partial V(\alpha)}{\partial\alpha}&=& -B\sin\alpha
- E\sin(\alpha+\bar\alpha) = 0\\[5pt]
\frac{\partial V(\bar\alpha)}{\partial\bar\alpha} 
&=& -D\sin\bar\alpha - E\sin(\alpha + \bar\alpha) = 0 
\end{array} \ .
\ee\\
%------------------------ 6  ---------------------------------------
\pagebreak

This gives the equations
$$
\sin\bar\alpha = 
\frac{B}{D}\sin\alpha\ \ \ \ 
$$
$$
B\sin\alpha+E(\sin\alpha\cos\bar\alpha+\sin\bar\alpha\cos\alpha)=0
$$

and the solutions are
$$
\cos\alpha=\frac{ED}{2}(\frac{1}{B^2}-\frac{1}{D^2}-\frac{1}{E^2})
$$
$$
\cos\bar\alpha=\frac{EB}{2}(\frac{1}{D^2}-\frac{1}{B^2}-\frac{1}{E^2}) .
$$
\\
We have clearly a minimum for a certain range of parameters, with non trivial
values of $\alpha$, $\bar\alpha$. This means that $CP$ is broken 
spontaneously.\\

$CP$ is broken at the high scale, it is transferred however to the Fermi scale via
the mixing of the Higgs representations which obey the restrictions (14).  
The $MSSM$ bi-doublets $h_u,h_d$ are then  (linear) combinations 
of the Higgs representations doublet components. The expressions involve quite a few
parameters, are very complicated and model dependent. The details are out of the
scope of this paper and I refer the reader to the papers ~\cite{Senjan}
~\cite{Fukuyama}.  
The only important relevant  fact for us is, that in all variants, the 
coefficients of those combinations involve $\Delta$ and $\bar\Delta$ 
(and a possibly complex
parameter $x$  that fixes the local symmetry breaking \cite{Senjan}) 
so that the $VEV$s\quad $<h_u>$,$<h_d>$ \quad are complex. \\

$H$ and $\overline\Sigma$ which come in the Yukawa coupling and contribute
to the mass matrices
$$M^i=Y_{10}^iH + Y_{\overline {126}}^i\overline\Sigma\\
$$
are given in 
terms of the physical $h_{u,d}$ as follows (the heavy combinations decouple):
\be
\begin{array}{cccccccc}  
H_{u,d}&=&a_u h_u &+& a_d h_d &+&\cdots&{\hbox{\small decoupled}}\\\
\bar\Sigma_{u,d}&=&b_u h_u& +& b_d h_d &+&\cdots& 
{\hbox{\small decoupled}}\\
\end{array}
\ee

The mass matrices are expressed then in terms of $<h_{u,d}>$
\be
M_u=(a_uY_{10}+b_uY_{\overline{126}})<h_u>
\ee
\be
M_d=(a_dY_{10}+b_dY_{\overline{126}})<h_d>
\ee
\be
M_\ell=(a_dY_{10}-3b_dY_{\overline{126}})<h_d>
\ee
\be
M_{\nu}^D=(a_uY_{10}-3b_uY_{\overline{126}})<h_u>
\ee
\be
M_{\nu_R}=Y_{\overline{126}}\bar\Delta 
\ee
\\
The mass matrices of the quarks and also leptons are therefore complex and
lead to a complex $CKM$ matrix as well as a complex $PNMS$ leptonic one.\\

\pagebreak
%------------------------------ 7 --------------------------

What was presented in this paper is only a possible realization of the 
scenario.
\\
To have a complete model, the free parameters must be fixed by fitting to the
 experimental data. For the minimal renormalizable SUSY SO(10), it was observed
~\cite{Senjan}\cite{Fukuyama}\cite{fit} that when $CP$ violation as well as the soft SUSY breaking terms 
are disregarded the model cannot be fully realistic.
The main difficulty lies in the fact that to get the right absolute masses of
the neutrinos one needs an intermediate symmetry breaking scale. This may
cause problems in particular for the gauge coupling unification.
Recently suggested solutions involve adding the $D({\bf 120})$ Higgs
representation \cite{120}, adding  type II seesaw  \cite{II},
considering possible contribution from soft SUSY breaking terms \cite{soft} 
or adding warped extra dimensions \cite{extra}. 
Our scenario is applicable in those cases also. 
It requires additional parameters and the superpotential is more complicated, 
yet the conjecture (15) leads to $SCPV$.\\

Recently, Grimus and K\"uhb\"ock \cite{grimus} \footnote{See also Aulakh
and Garg \cite{aulakh}} were able, by adding $D({\bf 120})$, to fit correctly
the fermionic masses and mixing, including the $CKM$ phase. Using a ${\bf Z_2}$-symmetry and specific requirements they reduced the number of free parameters.
They assumed also that the Yukawa couplings are real but did not explain how 
the complex VEVs are spontaneously generated. Applying  here our scenario
one can explicitly relate the high scale $CP$ violation to the $CKM$ one 
\cite{preparation}. 

\hspace{6 cm}

What about {\em the strong $CP$ problem}? \\

To solve the $QCD$ $\Theta$ problem in the framework of $SCPV$ one must must
add not only extra symmetries but also exotic fermions
\cite{Nelson}\cite{Barr}\cite{branco}, hence, to go beyond $SO(10)$.
The simplest solution, in the framework of the renormalizable $SO(10)$, is to 
require global $U(1)_{PQ}$~\cite{PQ} invariance with the invisible axion 
scenario~\cite{axion}~\cite{Langacker}. 
It is interesting then to observe that the energy range of our $SCPV$ lies 
within the invisible axion window~\cite{window}
\be
10^9 GeV \stackrel{<}{\sim} f_a \stackrel{<}{\sim}
10^{12} GeV\ ,
\ee
where $f_a$ is the axion decay constant.\\

This can  be applied to $SUSY SO(10)$ as well.
The minimal renormalizable $SUSY SO(10)\times U(1)_{PQ}$ was discussed 
recently in a paper by Fukuyama and Kikuchi~\cite{F+K}. The requirement
of $U(1)_{PQ}$ invariance using the $PQ$ charges
$$
PQ(\Psi)=-1,\quad PQ(H)=2, 
$$
$$
PQ(\Sigma)=-2,\quad PQ(\overline\Sigma)=2,\quad PQ(\Phi)=0
$$
forbids only two terms in the superpotential
\be
\begin{array}{ccl}
W_{PQ}&=&M\Phi^2 + \lambda_\Phi \Phi^3 + M_\Sigma \Sigma\overline\Sigma
+ \lambda_\Sigma \Phi\Sigma\overline\Sigma\\
&+& {\cal K}\Phi\Sigma H + 
\Psi_i(Y_{\scriptscriptstyle\bf 10}^{ij}H 
+ Y_{\overline{\scriptscriptstyle\bf 126}}^{ij} \overline\Sigma)\Psi_j\quad . 
\end{array}
\ee 
Hence, our scenario for $SCPV$ is still intact (although with different 
phases).\\

The breaking of local 
$B-L$ via the $VEV$s of $\overline\Sigma(\overline{126})$ and 
$\Sigma(126)$ will  
also break spontaneously the global $U(1)_{PQ}$ and explain the coincidence of 
the scales of the axion window and the seesaw one. In our scenario it will
also coincide with the scale of $SCPV$ and that of leptogenesis.\\

%Fukuyama and Kikuchi~\cite{F+K} suggest in their paper that the difference
%between the phases of $\Delta$ and $\bar\Delta$ is related to the
%axion\footnote{G.Senjanovi\'c claims however that it is not possible to break %two symmetries using one $VEV$ (private communication after my talk in Paris).%}.\\

\hspace{10.0cm}

{\Large Conclusions}\\

This paper is a version for publication of ref. \cite{pap}. I presented  
a scenario for $SCPV$ in both non-$SUSY$ and $SUSY \ SO(10)$. 
$CP$ is broken spontaneously at the scale of the $RH$ neutrinos but a phase
is generated also in the $CKM$ low energy mixing matrix. We have therefore
$CP$ violation at low and high energies as is required experimentally.\\
To the best of my knowledge there are no SUSY-GUT models that really 
discuss the way the phases are generated spontaneously. $SCPV$ is induced
in most models by giving ad-hoc phases by hand to some $VEV$s.\\
If $U(1)_{PQ}$ invariance is also used, one finds the interesting situation
that the scales of Baryogenesis, Seesaw, $SCPV$ and the breaking of 
$U(1)_{PQ}$ are all at the same order of magnitude.\\ 

%\hspace{20.0cm}
%--------------------------  8 ---------------------------------
\pagebreak

{\Large Appendix: the parameters of the scalar potential}\\

$$
\frac{\partial W_\Delta}{\partial \phi^u},\ \   \frac{\partial W_\Delta}
{\partial \phi^d},\ \   \frac{\partial W_\Delta}{\partial \phi_1},\ \
\frac{\partial W_\Delta}{\partial \phi_3},\ \   \frac{\partial W_\Delta}
{\partial H^u},\ \   \frac{\partial W_\Delta}{\partial H^d}
$$
do not give terms with a phase.\\

$\alpha$ dependent terms are obtained from\quad $ \frac{\partial W_\Delta}
{\partial \bar\Delta}$\quad and \quad $\frac{\partial W_\Delta}
{\partial \bar\Delta^u}$ 
\quad i.e.\\
$$
\left|\frac{\partial W_\Delta}{\partial \bar\Delta}\right|^2+
\left|\frac{\partial W_\Delta}{\partial \bar\Delta^u}\right|^2=constant +
B\cos\alpha
$$

Therefore,
$$
B=2\sigma\phi^u[M_\Sigma
+\frac{\lambda_\Sigma}{10}(\frac{1}{\sqrt6}\phi_1+\frac{1}{\sqrt2}\phi_2 +
\phi_3)][\frac{\lambda_\Sigma}{10}\Delta^d - \frac{\bar\kappa}{\sqrt5}H^d]
+\frac{\sqrt2}{75}\sigma\lambda_\Sigma^2\phi^d\phi_2\Delta^d=
$$
$B(M_\Sigma, \lambda_\Sigma, \bar\kappa, \phi_i,\phi^u,\phi^d,
\Delta^d,H^d )$ .\\

In the same way
$$
D=2\sigma\phi^d[M_\Sigma
+\frac{\lambda_\Sigma}{10}(\frac{1}{\sqrt6}\phi_1+\frac{1}{\sqrt2}\phi_2 +
\phi_3)][\frac{\lambda_\Sigma}{10}\Delta^u - \frac{\bar\kappa}{\sqrt5}H^u]
+\frac{\sqrt2}{75}\sigma\lambda_\Sigma^2\phi^u\phi_2\Delta^u=$$
$D(M_\Sigma, \lambda_\Sigma,\kappa, \phi_i,\phi^u,\phi^d,
\Delta^u,H^u) $ .\\

A term proportional to\quad $\cos(\alpha+\bar\alpha)$\quad is generated only by
$\frac{\partial W_\Delta}{\partial \phi_2}$.\\
Hence,
$$
E=\frac{1}{75}\lambda_\Sigma^2\bar\Delta^u\Delta^d\sigma^2=
E(\lambda_\Sigma,\sigma,\bar\Delta^u,\Delta^d)\ .
$$\\

\pagebreak
%-------------------------------- 9  ---------------------------------

\end{document}